\documentclass[showkeys,showpacs,prd]{revtex4}
\usepackage{amsmath}
\usepackage{amssymb}
\usepackage{graphicx}
\usepackage{yfonts}
\usepackage{color}
\usepackage{ulem}
\usepackage{tikz}

\begin{document}

\title{
Nonperturbative quantization \`{a} la Heisenberg for non-Abelian gauge theories: two-equation approximation
}

\author{
Vladimir Dzhunushaliev
}
\email{v.dzhunushaliev@gmail.com}
\affiliation{
	Dept. Theor. and Nucl. Phys., KazNU, Almaty, 050040, Kazakhstan \\
	IETP, Al-Farabi KazNU, Almaty, 050040, Kazakhstan \\
	Institut f\"ur Physik, Universit\"at Oldenburg, Postfach 2503,
	D-26111 Oldenburg, Germany \\
	Institute of Systems Science,
	Durban University of Technology, P. O. Box 1334, Durban 4000, South Africa
}

\begin{abstract}
The nonperturbative quantization technique \`{a} la Heisenberg is applied for non-Abelian  gauge theories.
The operator Yang-Mills equation is written, which on the corresponding averaging gives an infinite set of equations  for all Green functions. We split all degrees of freedom into two groups: in the former,  we have $A^a_\mu \in \mathcal G \subset SU(N)$, and in the second group we have coset degrees of freedom $SU(N) / \mathcal G$. Using such splitting and some assumptions about 2- and 4-point Green functions, we truncate the infinite set of equations to two equations. The first equation is for the gauge fields
from the subgroup $\mathcal G$, and the second equation is for a gluon condensate which is the dispersion of quantum fluctuations of the coset fields. Two examples are considered: The first one is a flux tube solution  describing longitudinal color electric fields stretched between quark and antiquark located at the $\pm$ infinities. The second one is a flux tube stretched between two quarks (antiquarks) located at 
$\pm \infty$. A special case is considered when the longitudinal electric field produced by a quark located at $+ \infty$ is equal and oppositely directed to the field generated by a quark located at $- \infty$ that leads to zero total electric field. Both solutions represents the dual Meissner effect: the electric field is pushed out from the gluon condensate.
\end{abstract}

\date{\today}

\maketitle

\section{Introduction}

In the 1950's, W. Heisenberg has offered the procedure of nonperturbative (NP) quantization for a nonlinear spinor field \cite{heis}. His purpose was to obtain all physical properties of electron from the first principles, i.e. from a fundamental equation, which he suggested to be the equation for a nonlinear spinor field. Following this approach, he was able to obtain, to some accuracy, the main properties of  electron.

Heisenberg's main idea was to write the operator nonlinear Dirac equation. Then, on multiplying  by field operators and performing subsequent quantum averaging,
one can obtain an infinite set of equations  for all Green functions.
In order to make practical use of  the infinite set of equations, he proposed to cut off this system of equations to obtain a finite set of equations.

In principle, this idea can be used for any strongly interacting fields. Here we employ this approach for quantum chromodynamics (QCD). 
Our main goals  are: (a) starting from the Yang-Mills operator equation, we write an infinite set of equations  for all Green functions; 
(b) using some assumptions, we truncate this set of equations  up to two equations describing gauge fields from the subgroup $\mathcal G \subset SU(N)$ 
and gluon condensate for gauge fields belonging to the coset $SU(N)/\mathcal G$; (c) we apply these two equation to obtain a flux tube solution describing the 
field distribution between quark and antiquark.

Similar approach for the perturbative calculations of field correlators can be found in Refs. \cite{Simonov:2009nf}-\cite{DelDebbio:1994zn} 
where the authors consider the formalism of gauge-invariant nonlocal correlators in non-Abelian gauge theories and derive nonlinear equations for field correlators in the large $N_c$ gluodynamics.

One of unsolved problems in quantum chromodynamics (QCD) is the problem of field distribution between quark and antiquark. Similar problem is easily solved in electrodynamics: the distribution of electric field between positive and negative charges can be easily found since Maxwell's electrodynamics is a linear theory. In QCD the problem is that the calculations should be done for non-perturbatively quantized fields because  Yang-Mills theories are strongly nonlinear ones. The standard point of view 
is that in QCD there is the dual Meissner effect: longitudinal electric field lines get compressed to a flux tube.

The flux tube field distribution is investigated within the framework of lattice QCD. In Ref. \cite{Ichie:2002dy}, the Abelian color flux of two- and 
three-quark systems in the maximally Abelian gauge in lattice QCD with dynamical fermions is investigated. In Refs.~\cite{Shibata:2014tpa} and \cite{Shibata:2012ae}, 
the non-Abelian dual Meissner effect in the SU(3) Yang-Mills theory is investigated by measuring the chromoelectric flux created by an quark-antiquark source. 
Lattice calculations strongly support the idea of the dual Meissner effect in QCD. However, for a more complete understanding of the nature of confinement, 
it is necessary to have at least approximate analytical calculations confirming this point of view.

Here we will show that applying the non-perturbative quantization \`{a} la Heisenberg for QCD and using the two-equation approximation, a solution describing the flux tube between quark and antiquark located at $\pm \infty$ can be obtained. The solution is characterized by a longitudinal color electric field directed from quark to antiquark. All fields in this solution are expelled by a condensate of coset gauge fields into the flux tube. This is a non-Abelian version of the field distribution between positive and negative charges in Maxwell's electrodynamics. Then we want to consider a non-Abelian version of the field distribution between charges with the same sign in Maxwell's electrodynamics. We expect that in this case we will have two longitudinal electric fields directed oppositely. We will consider some special case when these fields are the same that leads to zero longitudinal color electric field in the flux tube.

\section{Nonperturbative  quantization \`{a} la Heisenberg for non-Abelian gauge theories}

Following Heisenberg, we write the SU(N) Yang-Mills equations as operator equations
\begin{equation}\label{1-10}
	D_\nu \widehat {F}^{a \mu\nu} = 0,
\end{equation}
where
$\hat F^B_{\mu \nu} = \partial_\mu \hat A^B_\nu - \partial_\nu \hat A^B_\mu +
g f^{BCD} \hat A^C_\mu \hat A^D_\nu$ is the field strength operator;
$\hat A^B_\mu$ is the gauge potential operator; $B, C, D = 1, \ldots , N$ are the SU(N) color indices; $g$ is the coupling constant; $f^{BCD}$ are the structure constants for the SU(N) gauge group.

How we can solve this equation? Following Heisenberg \cite{heis}, we have to write an infinite set of equations  for all Green functions
\begin{eqnarray}
	\left\langle
		D_\nu \widehat {F}^{A \mu\nu} (x)
	\right\rangle &=& 0 ,
\label{1-20}\\
	\left\langle
		\hat A^{B_1}_{\alpha_1} (x_1)
		D_\nu \widehat {F}^{A \mu\nu} (x)
	\right\rangle &=& 0 ,
\label{1-30}\\
	\left\langle
		\hat A^{B_1}_{\alpha_1} (x_1) \hat A^{B_2}_{\alpha_2} (x_2)
		D_\nu \widehat {F}^{A \mu\nu} (x)
	\right\rangle &=& 0 ,
\label{1-40}\\
	\ldots &=& 0	,
\label{1-50}\\
	\left\langle
		\hat A^{B_1}_{\alpha_1} (x_1) \ldots \hat A^{B_n}_{\alpha_n} (x_n)
		D_\nu \widehat {F}^{A \mu\nu} (x)
	\right\rangle &=& 0
\label{1-60}\\
	\ldots &=& 0	.
\label{1-70}
\end{eqnarray}
Here
$
\left\langle (\ldots) \right\rangle =
\left\langle Q \left| (\ldots) \right| Q \right\rangle
$ and $\left. \left. \right| Q \right\rangle$ is a quantum state of the given physical system. For  brevity, we will write $\left. \right\rangle$ instead of
$\left. \left. \right| Q \right\rangle$.

The first equation \eqref{1-20} is the equation for the 1-st order Green function $G^B_{\mu} (x) = \left\langle A^B_{\mu}(x) \right\rangle$.
But it contains Green functions of the 2-nd order $\left\langle A \partial A \right\rangle$ and 3-rd order $\left\langle A^3 \right\rangle$ (for brevity, we omit all indices).
The second equation \eqref{1-30} is the equation for the 2-nd order Green function
$G^{BC}_{\mu \nu} (x,x) = \left\langle A^B_{\mu}(x) A^c_{\nu}(x) \right\rangle$.
But it contains Green functions of the 3-rd order $\left\langle A^2 \partial A \right\rangle$ and 4-th order $\left\langle A^4 \right\rangle$.
We see that such situation holds  true for all other equations \eqref{1-40}-\eqref{1-70}. Thus, to close this set of equations, we have to write an infinite system of equations.

The solution of the full set of equations  \eqref{1-20}-\eqref{1-70} gives us full information on a quantum state $\left. \left. \right| Q \right\rangle$ and field operators $\hat A^B_\mu$.
In this sense, we can say that the solution of the full set of equations \eqref{1-20}-\eqref{1-70} is the solution of the operator field equations \eqref{1-10}.

We would like to note that in Ref. \cite{Frasca:2015yva} the author considers Dyson - Schwinger equations set  \eqref{1-20}-\eqref{1-70} for the Yang-Mills theory stopping to the two-point function. We have to note that in our approach presented here we use some approximation in order to work with 4 - point Green function. 

\section{Two-equation approximation}

In practice,  we cannot solve the infinite set of equations  \eqref{1-20}-\eqref{1-70}. We need to cut it off  to obtain a \emph{finite} set of equations.
Then the solution of the truncated set of equations  \eqref{1-20}-\eqref{1-70} will approximately describe the solution of the full system.
In order to do so, we have to use some physical intuition. For example, we can assume that $n$-th Green function is a polylinear combination of $m < n$
Green functions and use it either to cut off the set of equations  \eqref{1-20}-\eqref{1-70} or to write an effective Lagrangian which will be an averaged $SU(N)$ Lagrangian.
Probably there exist other variants to cut off the infinite set of equations  \eqref{1-20}-\eqref{1-70}.

In this section we use the following strategy:  we assume that in some physical situations all $SU(N)$ degrees of freedom can be decomposed into two groups.
In the first group, the gauge fields
$\hat A^a_\mu = \langle \hat A^a_\mu \rangle + i \delta \hat A^a_\mu \in \mathcal G \subset SU(N)$, where $\langle \hat A^a_\mu \rangle = A^a_\mu$
will be treated  as classical fields, and
$\delta \hat A^a_\mu$ is the quantum fluctuation around the classical field $A^a_\mu$. In the second group, the gauge fields $A^m_\mu \in SU(N) / \mathcal G$
are pure quantum ones in the sense that
$\langle \hat A^m_\mu \rangle = 0$.

We will consider  physical systems where the quantum average of odd degrees of the gauge field are zero,
\begin{equation}\label{1-80}
	\left\langle
      \left( \hat A^{m_1}_{\mu_1}(x_1) \ldots \hat A^{m_{2k+1}}_{\mu_{2k+1}}(x_{2k+1}) \right)
	\right\rangle = 
	\left\langle
	\left( \delta \hat A^{a_1}_{\mu_1}(x_1) \ldots 
	\delta \hat A^{a_{2k+1}}_{\mu_{2k+1}}(x_{2k+1}) \right)
	\right\rangle = 
	0 .
\end{equation}
The decomposition of field strengths $\hat F^a_{\mu \nu}$ and $\hat F^m_{\mu \nu}$ into $(\cdots)^a$ and $(\cdots)^m$ parts can be written as follows:
\begin{eqnarray}
	\hat F^a_{\mu \nu} &=& \hat{\mathcal F}^a_{\mu \nu} +
	g f^{amn} \hat A^m_\mu \hat A^n_\nu ,
	\label{1-82}\\
	\hat F^m_{\mu \nu} &=& \partial_{[ \mu} \hat A^m_{\nu ]} -
	g f^{amn} \hat A^a_{[ \mu} \hat A^n_{\nu ]} +
	g f^{mpq} \hat A^p_\mu \hat A^q_\nu.
	\label{1-84}
\end{eqnarray}
Here $
\hat{\mathcal F}^a_{\mu \nu} = \partial_\mu \hat A^a_\nu -
\partial_\nu \hat A^a_\mu + g f^{abc} \hat A^b_\mu \hat A^c_\nu
$ is the field strength tensor of the subgroup $\mathcal G$;
$
\hat A^B_{[\mu} \hat A^C_{\nu ]} ] = \hat A^B_\mu \hat A^C_\nu - \hat A^B_\nu \hat A^C_\mu
$ is the antisymmetrization procedure; $a, b, c, \ldots$ are the subgroup indices $\mathcal G$, and $m, n, p, \dots$ are the coset indices.
We will consider such subgroup $\mathcal G$ in order to have structure constants $f^{mpq} = 0$, where the indices $m, p, q$ are in the coset $SU(N) / \mathcal G$.
For example, it can be pairs $U(1) \subset SU(2)$ or $SU(2) \times U(1) \subset SU(3)$.

Let us consider the equation \eqref{1-20} for $A = a$. After algebraic manipulations, we have
\begin{equation}\label{1-90}
	\tilde D_\nu \mathcal F^{a \mu \nu} - \left( m^2 \right)^{a b \mu \nu} A^b_\nu
    + \left( \mu^2 \right)^{ab \mu\nu} A^b_\nu = j^{a \mu},
\end{equation}
where $
\tilde D_\mu = \partial_\mu + g f^{abc} A^b_\mu$ is the covariant derivative in the subgroup
$\mathcal G$ (for details see Appendix \ref{averaged}).

2-point Green functions for the gauge fields $\delta \hat A^a_\mu \in \mathcal G$ and for the coset  $\hat A^m_\mu \in SU(N) /\mathcal G$ are defined as
\begin{eqnarray}
	G^{mn \mu \nu}(y,x) &=& \left\langle
        \hat A^{m \mu}(y) \hat A^{n \nu}(x)
    \right\rangle,
\label{1-120}\\
	G^{ab \mu \nu}(y,x) &=& \left\langle
	\delta \hat A^{a \mu}(y) \delta \hat A^{b \nu}(x)
	\right\rangle .
\label{1-125}
\end{eqnarray}
Equation \eqref{1-20} for $A = m$ gives us
\begin{equation}\label{1-130}
	\left\langle
	   D_\nu \widehat {F}^{m \mu\nu}
	\right\rangle = 0.
\end{equation}
Here we took into account the condition \eqref{1-80} and the simplification that  quantum fields from the subgroup
$\mathcal G$ and the coset $SU(n) / \mathcal G$ do not correlate,
\begin{equation}\label{1-135}
    \left\langle
       \hat A^m_\nu \delta \hat A^a_\mu
    \right\rangle = 0 .
\end{equation}
Next step is a consideration of equation \eqref{1-30} for $A = m, B_1 = r, \alpha_1 = \alpha$. After tedious calculations, we have (for details see Appendix \ref{averaged})
\begin{equation}
\begin{split}
	&
	\left[
	\partial_{x^\nu} \partial^{x^\mu} G^{rm \alpha \nu} (y, x) -
	\Box_x G^{rm \alpha \mu} (y, x)
	\right]_{y = x} +
\\
	&
	g f^{amn} \biggl \{
	- \partial_{x^\nu} \left[
	A^{a \mu} (x) G^{rn \alpha \nu} (y, x) -
	A^{a \nu} (x) G^{rn \alpha \mu} (y, x)
	\right]_{y = x} -
	A^a_\nu (x) \left[
	\partial^{x^\mu} G^{rn \alpha \nu} (y, x) -
	\partial^{x^\nu} G^{rn \alpha \mu} (y, x)
	\right]_{y = x} +
\\
	&		
	G^{rn \alpha}_{\phantom{rn \alpha} \nu} (x, x)
	\mathcal F^{a \mu \nu}(x)
	\biggl \} +
	g^2 f^{amn} f^{bnp}  A^a_\nu (x)
	\biggl[
	A^{b \mu} (x) G^{rp \alpha \nu} (x, x) - A^{b \nu} (x) G^{rp \alpha \mu} (x, x)
	\biggl] +
\\
	&
	g^2 f^{amn} f^{apq} G^{rnpq \alpha \phantom{\nu} \mu \nu}_{\phantom{rnpq \alpha} \nu} (x,x,x,x)
	= 0.
\label{1-140}
\end{split}
\end{equation}
Here we have introduced a 4-point Green function
\begin{equation}\label{1-150}
	G^{mnpq}_{\phantom{mnpq}\mu \nu \rho \sigma}(x, y, z, u) = \left\langle
    \hat A^m_\mu(x) \hat A^n_\nu(y) \hat A^p_\rho(z) \hat A^q_\sigma(u)
  \right\rangle .
\end{equation}
In order to close the equations \eqref{1-90} and \eqref{1-140}, we have to define the last term in \eqref{1-140}: 4-point Green function \eqref{1-150}.

Finally, two-equation approximation of the NP quantization \`{a} la Heisenberg is (see the equations \eqref{a-120} and \eqref{a-130} from Appendix \ref{averaged})
\begin{eqnarray}
	&&
	\tilde D_\nu \mathcal F^{a \mu \nu} - \left[
	\left( m^2 \right)^{ab \mu \nu} -
	\left( \mu^2 \right)^{ab \mu \nu}
	\right] A^b_\nu = j^{a \mu} ,
\label{1-160} \\
	&&
	\left[
	\partial_{x^\nu} \partial^{x^\mu} G^{rm \alpha \nu} (y, x) -
	\Box_x G^{rm \alpha \mu} (y, x)
	\right]_{y = x} +
\nonumber \\
&&
	g f^{amn} \biggl \{
	- \partial_{x^\nu} \left[
	A^{a \mu} (x) G^{rn \alpha \nu} (y, x) -
	A^{a \nu} (x) G^{rn \alpha \mu} (y, x)
	\right]_{y = x} -
	A^a_\nu (x) \left[
	\partial^{x^\mu} G^{rn \alpha \nu} (y, x) -
	\partial^{x^\nu} G^{rn \alpha \mu} (y, x)
	\right]_{y = x} +
\nonumber \\
&&		
	G^{rn \alpha}_{\phantom{rn \alpha} \nu} (x, x)
	\mathcal F^{a \mu \nu}(x)
	\biggl \} +
	g^2 f^{amn} f^{bnp}  A^a_\nu (x)
	\biggl[
	A^{b \mu} (x) G^{rp \alpha \nu} (x, x) - A^{b \nu} (x) G^{rp \alpha \mu} (x, x)
	\biggl] +
\nonumber \\
	&&
	g^2 f^{amn} f^{apq} G^{rnpq \alpha \phantom{\nu} \mu \nu}_{\phantom{rnpq \alpha} \nu} (x,x,x,x)
	= 0 .
\label{1-170}
\end{eqnarray}
The first equation \eqref{1-160} describes the dynamics of $\mathcal G$ gauge fields interacting with a coset condensate of quantum degrees of freedom belonging to the coset $SU(N) / \mathcal G$.
The condensate is described by 2- and 4-point Green functions and it is defined by the second equation \eqref{1-170}.
In order to have a closed set of equations, we have to make some assumptions about a 4-point Green function. For example, it can be expressed as a bilinear combination of 2-point Green functions $G^{(2)}$:
\begin{equation}\label{1-180}
    G^{(4)} \approx
    C_1 \left(G^{(2)} \right)^2 + C_2 G^{(2)} + C_3,
\end{equation}
where $C_{1,2,3}$ are constants (here we omit all indices). Probably the simplest example is a Mexican hat approximation that schematically looks like
\begin{eqnarray}
    G^{(4)} \approx \frac{\lambda}{4} \left(
        G^{(2)} - m^2  \
    \right)^2 + \text{const}.
\label{1-185}
\end{eqnarray}
Similar decomposition is used in Ref. \cite{Calzetta:1999xh} to derive the Boltzmann - Langevin equation as the correct description of the kinetic limit of quantum field theory.

\subsection{Qualitative consideration of possible regular solutions}

Here we want to consider possible solutions of the equations \eqref{1-160} and \eqref{1-170}.
Taking into account the expressions for the mass matrix $\left( \mu^2 \right)^{ab\mu\nu}$ and the current $j^{a\mu}$, we can assume that regular solutions with
\begin{equation}\label{2a-10}
	G^{mp \mu \nu}(x,x) \stackrel{r, \rho \rightarrow \infty}{\longrightarrow} m^2
\end{equation}
may exist, where $r$ is for the spherical case, $\rho$ is for the cylindrical case, and $m^2 = \text{const}$. Then the classical field $A^a_\mu$ will decay as
\begin{eqnarray}\label{2a-20}
	A^a_\mu &\propto& \frac{e^{- m r}}{r}, \text{ for the spherical case} ,
\\
	A^a_\mu &\propto& \frac{e^{- m \rho}}{\sqrt \rho}, \text{ for the cylindrical case},
\label{2a-25}
\end{eqnarray}
and the equation \eqref{1-170} is also satisfied since
$\partial_\mu G^{(2)} \rightarrow 0$ and $G^{(4)} \rightarrow 0$ by virtue of \eqref{1-180}.

Physically such solutions will present either a hedgehog filled with the classical non-Abelian
$\mathcal G$ gauge field or a flux tube. The gluon condensate created by the coset gauge fields
$A^m_\mu \in SU(N) / \mathcal G$ confines classical $\mathcal G$ gauge fields either in the ball or in the tube.

\subsection{Energy density}

In this section we want to calculate the energy density for the two-equation approximation:
\begin{equation}
	\varepsilon = \left \langle
	\frac{\left( E^A_{i} \right )^2 + \left( H^A_{i} \right )^2}{8 \pi}
	\right \rangle,
	\label{2-10}
\end{equation}
where $E^A_i = F^A_{0i}$ is the color electric field; $H^A_i = \frac{1}{2} \epsilon_{ijk} F^A_{jk}$ is the color magnetic field,
and $\epsilon_{ijk}$ is the Levi-Civita symbol. Taking into account \eqref{1-82}-\eqref{1-84} and the following algebraic properties of the subgroup $\mathcal G \subset SU(N)$
\begin{equation}
	f^{abc} , f^{amn } \neq 0 , f^{mnp} = 0, \quad a,b,c \in \mathcal G ;
	m, n, p \in SU(N) / \mathcal G,
	\label{2-20}
\end{equation}
we obtain
\begin{equation}
	\begin{split}
		8 \pi \varepsilon = & \left\langle 
			\left (\mathcal E^a_i \right)^2 
		\right\rangle + 
		\left\langle 
			\left (\mathcal H^a_i \right)^2 
		\right\rangle - 2 g f^{amn} 
		\left\langle \mathcal E^a_{i} \right\rangle G^{mn 0i}(x,x) +
		2 g f^{amn} \epsilon_{ijk} \left\langle \mathcal H^a_{k} \right\rangle G^{mn ij}(x,x) -
\\
		&
		\left\langle 
			\partial_{[0} A^m_{i]} \partial^{[0} A^{(m)i}
		\right\rangle +
		\left\langle 
			\partial_{[i} A^m_{j]} \partial^{[i} A^{(m)j}
		\right\rangle +
\\
		&
		g^2 \left\{
			- f^{amn} f^{apq} G^{mnpq \phantom{0i} 0i}_{\phantom{mnpq} 0i}(x,x,x,x) +
			f^{amn} f^{bmq} \left[ 
			- \left\langle 
				A^a_{[ 0} A^n_{i ]} A^{b [ 0} A^{(q) i]}
			\right\rangle  +
			\left\langle 
				A^a_{[ i} A^n_{j ]} A^{b [ i} A^{(q) j]}
			\right\rangle
		\right]
		\right\} 
	\end{split}
	\label{2-130}
\end{equation}
here we take into account $f^{amn} G^{mn}_{\phantom{mn}\mu \nu} = 0$. The interesting thing here is that, in principle, one can obtain regular solutions with finite energy even if the 2-point Green function $G^{mn}_{\phantom{mn} \mu \nu}$ is a nonzero constant at  infinity. In such case it is necessary that the 4-point Green function $G^{mnpq}_{\phantom{mnpq} \alpha \beta \gamma \delta}$ and color electric $\mathcal E^a_i$ and magnetic $\mathcal H^a_i$ fields tend to zero at  infinity. It can happen if we choose the bilinear decomposition \eqref{1-180}. In this case, even if $G^{(2)} \rightarrow m^2$,  the 4-point Green function $G^{(4)} \rightarrow 0$, and if in addition the color electric and magnetic fields tend to zero, we will have the solution with finite energy. The equations \eqref{a-40} and \eqref{a-132} show that at  infinity the mass matrix goes to a constant, and the current $j^{a \mu} \rightarrow 0$. Then asymptotically the equation \eqref{1-160} gives us exponentially decreasing color gauge fields $\mathcal E^a_i$ and $\mathcal H^a_i$.

\section{Infinite SU(3) flux tube between quark and antiquark}

\begin{figure}[h]
	\begin{minipage}[ht]{.45\linewidth}
		\begin{center}
			\fbox{
				\includegraphics[width=.9\linewidth]{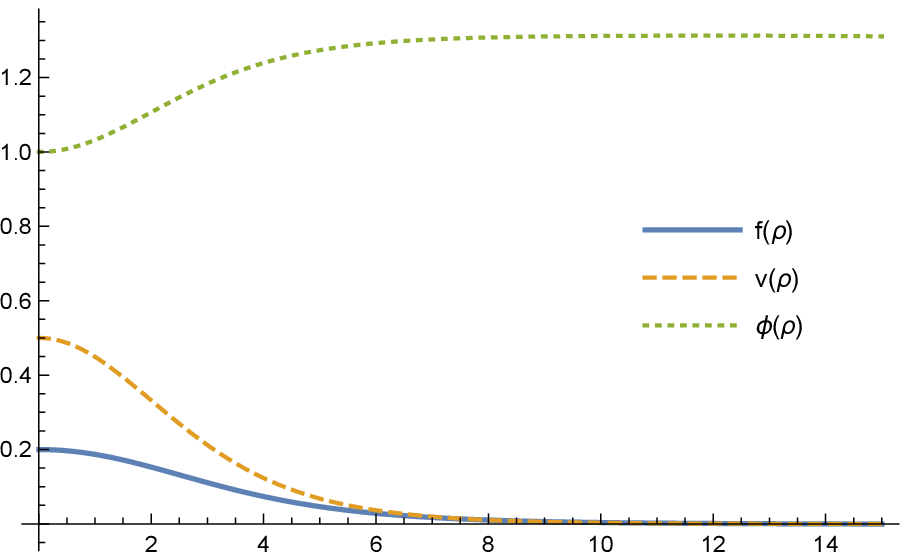}
			}
		\end{center}
		\caption{The functions $f(\rho), v(\rho), \tilde \phi(\rho)$.
			The solid curve is $f(\rho)$,
			the dashed curve  is $v(\rho)$,
			the dotted curve is $\tilde \phi(\rho)$.
			$\mu_1 = 1.2325683, \mu_2 = 1.180660003, M = 1.3137067$, $\tilde \lambda = 0.1$.
			}
		\label{potentials}
	\end{minipage}\hfill
	\begin{minipage}[ht]{.45\linewidth}
		\begin{center}
			\fbox{
				\includegraphics[width=.9\linewidth]{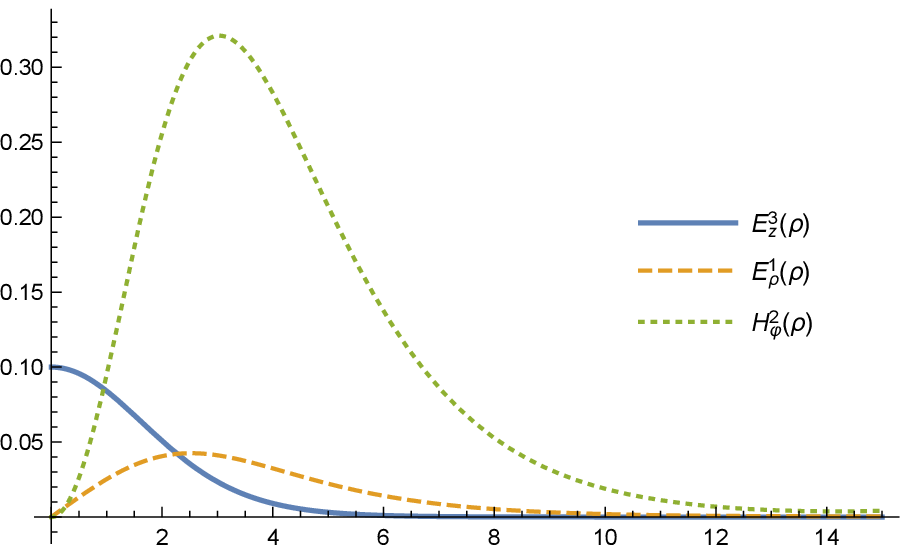}
			}
		\end{center}
		\caption{The chromoelectric and chromomagnetic fields
			$E^3_z(\rho), E^1_\rho(\rho), H^2_\varphi (\rho)$.
			The solid curve is $E^3_z(\rho)$,
			the dashed curve  is $E^1_\rho(\rho)$,
			the dotted curve is $H^2_\varphi (\rho)$
		}
		\label{fields}
	\end{minipage}
\end{figure}

In this section we would like to show that the set of equations  \eqref{c-10} and \eqref{3c-70}
\begin{eqnarray}
	\tilde D_\nu \mathcal F^{a \mu \nu} - \left[
	\left( m^2 \right)^{ab \mu \nu} -
	\left( \mu^2 \right)^{ab \mu \nu}
	\right] A^b_\nu &=& 0 ,
\label{4-4}\\
	\Box \phi - \left( m^2_\phi \right)^{ab \mu \nu} A^a_\nu A^b_\mu \phi -
	\lambda \phi \left( M^2 - \phi^2  \right) &=& 0
\label{4-8}
\end{eqnarray}
has a flux tube solution where a longitudinal chromoelectric field is pushed out from a gluon condensate in the flux tube;
$
\left( m^2 \right)^{ab \mu \nu} A^b_\nu
$,
$
\left( \mu^2 \right)^{ab \mu \nu} A^b_\nu
$ and
$
\left( m^2_\phi \right)^{ab \mu \nu} A^a_\nu A^b_\mu \phi
$ are calculated in Appendix \ref{adaptation}. We seek a solution in the form
\begin{equation}
	A^1_t(\rho) = \frac{f(\rho)}{g} ; \quad A^2_z(\rho) = \frac{v(\rho)}{g} ;
	\quad \phi(\rho) = \phi(\rho).
\label{4-10}
\end{equation}
Here we use the cylindrical coordinate system $t, z, \rho , \varphi$. The corresponding color electric and magnetic fields are then
\begin{eqnarray}
	E^3_z(\rho) &=& F^3_{tz} = \frac{f(\rho) v(\rho)}{g}  ,
\label{4-70}\\
	E^1_\rho(\rho) &=& F^1_{t \rho} = -\frac{f'(\rho)}{g} ,
\label{4-80}\\
	H^2_\varphi (\rho) &=& \rho \epsilon_{\varphi \rho z} F^{2\rho z} =
	- \rho \frac{v'(\rho)}{g}.
\label{4-90}
\end{eqnarray}
Substituting \eqref{4-10} into the equations \eqref{4-4} and \eqref{4-8}, we have
\begin{eqnarray}
	f'' + \frac{f'}{\rho} &=& f \left( v^2 + m^2 \phi^2 - \mu^2_1 \right),
\label{sec1-30}\\
	v'' + \frac{v'}{\rho} &=& v \left( - f^2 + m^2 \phi^2 - \mu^2_2 \right),
\label{4-20}\\
	\phi'' + \frac{\phi'}{\rho} &=& \phi \left[
	\frac{m^2_\phi}{g^2} \left( - f^2 + v^2 \right)
	+ \lambda \left( \phi^2 - M^2 \right)\right].
\label{4-30}
\end{eqnarray}
Here the prime  denotes  differentiation with respect to $\rho$. 
By redefining 
$\phi = \tilde \phi / m$, $M = \tilde M / m$, and $\lambda = m^2 m^2_\phi \tilde \lambda / g$, we have
\begin{eqnarray}
	f'' + \frac{f'}{\rho} &=& f \left( v^2 + \tilde \phi^2 - \mu^2_1 \right),
\label{4-40}\\
	v'' + \frac{v'}{\rho} &=& v \left( - f^2 + \tilde \phi^2 - \mu^2_2 \right),
\label{4-50}\\
	\tilde \phi'' + \frac{\tilde \phi'}{\rho} &=&
	\frac{m^2_\phi}{g^2} \tilde \phi \left[ - f^2 + v^2 
	+ \tilde \lambda \left( \tilde \phi^2 - \tilde M^2 \right)\right]
\label{4-60}
\end{eqnarray}
here according to \eqref{c-110} $m^2_\phi/g^2 = 1/4$. Equations \eqref{4-40}-\eqref{4-60} are solved as a nonlinear eigenvalue problem with the eigenvalues $\mu_{1,2}, \tilde M$ and the eigenfunctions
$f, v, \tilde \phi$. The boundary conditions are
\begin{equation}
\begin{split}
	f(0) &= 0.2, f'(0) = 0;
\\
	v(0) &= 0.5, v'(0) = 0;
\\
	\phi(0) &= 1.0, \phi'(0) = 0 .
\label{4-100}
\end{split}
\end{equation}
The results of calculations are presented in Figs. \ref{potentials} and  \ref{fields}.
Thus, we have obtained the flux tube stretched between two infinitely separated quark and antiquark.
This is the non-Abelian analogue of electric field distribution between positive and negative electric charges in the Maxwell electrodynamics. In contrast with Abelian electrodynamics, the electric field is confined within the tube due to the self-interaction of the non-Abelian fields.

From Eq's \eqref{sec1-30}-\eqref{4-60} we can obtain the asymptotic behavior of functions
$f(\rho), v(x)$ and $\tilde \phi(\rho)$ is
\begin{eqnarray}
	f(\rho) &\approx& f_0 \frac{e^{- \rho \sqrt{M^2 - \mu_1^2}}}{\sqrt{\rho}} ,
\label{4-110}\\
	v(\rho) &\approx& v_0 \frac{e^{- \rho \sqrt{M^2 - \mu_2^2}}}{\sqrt{\rho}} ,
\label{4-120}\\
	\tilde \phi(\rho) &\approx& M -
	\phi_0 \frac{e^{- \rho \sqrt{2\lambda M^2}}}{\sqrt{\rho}}
\label{4-130}
\end{eqnarray}
where $f_0, v_0$ and $\phi_0$ are some constants.
\par
\begin{figure}[h]
	\begin{center}
		\fbox{
			\includegraphics[width=10cm,height=7cm]{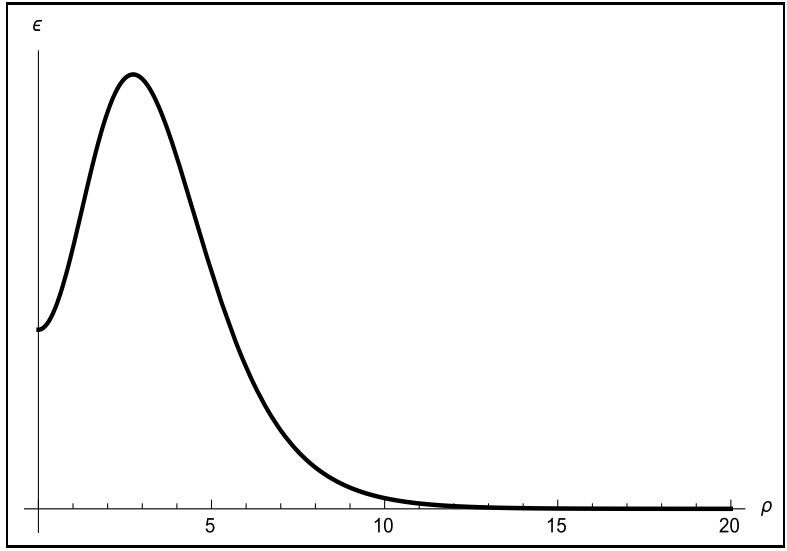}}
		\caption{The energy density $\epsilon(x)$.}
		\label{energy}
	\end{center}
\end{figure}

The linear energy density that follows from \eqref{2-130} and \eqref{4-10} - \eqref{4-90} is
\begin{equation}
	8 \pi \epsilon(\rho) = \frac{{f^\prime}^2}{2} + \frac{{v^\prime}^2}{2} + 
	\frac{{\phi^\prime}^2}{2} + \frac{f^2 v^2}{8} + m^2 \frac{f^2 \phi^2}{8} + 
	m^2 \frac{v^2 \phi^2}{8} + \frac{\mu_1^2}{2} f^2 - \frac{\mu_2^2}{2} v^2 + 
	\frac{\lambda}{4} \left( 
	\phi^2 - M^2
	\right)
\label{4-140}
\end{equation}
and is plotted in Fig. \ref{energy}. According to \eqref{4-110} - \eqref{4-130} the linear energy density exponentially decreases at the infinity.
\par
The flux of the longitudinal electric field is
\begin{equation}
	\Phi = \int E^3_z ds = \frac{2 \pi}{g}
	\int^{\infty}_{0} \rho f(\rho)v(\rho) d\rho < \infty.
\label{4-150}
\end{equation}

It is useful to list the properties of the physical system under consideration for which the flux solution is obtained:
\begin{itemize}
	\item the gauge fields belonging to the subgroup $\hat A^a_\mu \in SU(2) \times U(1) \subset SU(3)$ and the coset $\hat A^m_\mu \in SU(3) / (SU(2) \times U(1))$ have different dynamics;
	\item quantum expectation values $\left\langle \hat A^a_\mu \right\rangle \neq 0$, but
	$\left\langle \hat A^m_\mu \right\rangle = 0$;
	\item 2-point Green functions of the coset fields $G^{mn \mu \nu}$  are approximately proportional to the total dispersion $D$ ($D$ is defined in Appendix \ref{approximation}):
	$G^{mn \mu \nu}(x, x) \approx \delta^{mn} \mathcal A^\mu \mathcal A^\nu  D(x, x)$;
	\item 2-point Green functions of the subgroup fields $G^{ab \mu \nu}$  are approximately constant: $G^{ab \mu \nu}(x, x) \approx \Delta^{ab} \mathcal B^\mu \mathcal B^\nu$;
	\item $\hat A^a_\mu = A^a_\mu + i \delta \hat A^a_\mu$, and the dispersion defect
	$
	\left\langle \delta \hat A^a_\mu \delta \hat A^b_\nu \right\rangle
	$ gives rise to:
		\begin{itemize}
			\item the appearing of masses for the non-Abelian fields $A^a_\mu$;
			\item  that the masses
			$
			\left( \mu^2 \right)^{ab\mu\nu}
			$ have to have a necessary sign for getting the flux tube solution;
		\end{itemize}	
	\item  the constants $m^2$ and $m^2_\phi$ are defined by the dispersions $G^{mn\mu\nu}$ of quantum fluctuations;
	\item  the closure procedure for the 4-point Green function
	$G^{mnpq}_{\phantom{mnpq} \alpha \beta \mu \nu}$ which represents a 4-point Green function as a bilinear combination of the total dispersion:
	$
	G^{mnpq \alpha \beta \mu \nu}(x,x,x,x) \approx C^{mnpq \alpha \beta \mu \nu}
	D(x,x) \left[ M^2 - D(x,x) \right]
	$.
	\item in order to have different masses $\mu_1 \neq \mu_2$,  the color anisotropy for the quantum gauge fields $\delta \hat A^a_\mu \in SU(2)$ is necessary.
\end{itemize}
The flux tube solution obtained in this section shows us that in this situation we deal with the dual Meissner effect: the gluon condensate pushes out the color electric field. In this connection we want to note that in Ref. \cite{Baker:1998jw} the interaction between gluon condensate and dual Meissner effect is investigated.

\section{Infinite flux tube solution between quark and quark (antiquark and antiquark)}

Here we want to consider a non-Abelian version of the field distribution between charges with the same sign in Maxwell's electrodynamics. We expect that in this case we will have two longitudinal electric fields directed oppositely. We will consider some special case when these fields are the same that leads to zero longitudinal color electric field in the flux tube.

The set of equations describing such a situation is
\begin{eqnarray}
	\tilde D_\nu F^{a \mu \nu} - \left[
		\left( m^2 \right)^{ab \mu \nu} -
		\left( \mu^2 \right)^{ab \mu \nu}
	\right] A^b_\nu &=& 0 ,
\label{5-10}\\
	\Box \phi - \left( m^2_\phi \right)^{ab \mu \nu} A^a_\nu A^b_\mu \phi -
	\lambda \phi \left( M^2 - \phi^2  \right) &=& 0,
\label{5-20}
\end{eqnarray}
where
\begin{eqnarray}
	\left( m^2 \right)^{ab \mu\nu} &=& - g^2 \left[
		f^{abc} f^{cpq} G^{pq \mu\nu} -
		f^{amn} f^{bnp} \left(
			\eta^{\mu \nu} G^{mp \phantom{\alpha} \alpha}_{\phantom{mn} \alpha} -
			G^{mp \nu \mu}
		\right)
	\right] ,
\label{5-30} \\
	\left( \mu^2 \right)^{ab \mu \nu} &=& - g^2 \left(
		f^{abc} f^{cde} G^{de \mu \nu} +
		\eta^{\mu \nu} f^{adc} f^{cbe} G^{de \phantom{\alpha} \alpha}_{\phantom{de} \alpha} +
		f^{aec} f^{cdb} G^{ed \nu \mu}
	\right) ,
\label{5-40} \\
	\left( m^2_\phi \right)^{ab \mu \nu} &=&
	g^2 f^{amn} f^{bnp} \frac{
		G^{mp \mu \nu} - \eta^{\mu \nu} G^{mp \alpha}_{\phantom{mp \alpha} \alpha}
	}{G^{mm \alpha}_{\phantom{mm\alpha} \alpha}} .
\label{5-60}
\end{eqnarray}
2-point Green functions for the gauge fields $\delta \hat A^a_\mu \in SU(2) \times U(1)$ and for the coset  $\hat A^m_\mu \in SU(3) / (SU(2) \times U(1))$ are defined as
\begin{eqnarray}
	G^{mn \mu \nu}(y,x) &=& \left\langle
		\hat A^{m \mu}(y) \hat A^{n \nu}(x)
	\right\rangle ,
\label{5-70}\\
	G^{ab \mu \nu}(y,x) &=& \left\langle
		\delta \hat A^{a \mu}(y) \delta \hat A^{b \nu}(x)
	\right\rangle ,
\label{5-80}
\end{eqnarray}
where
$F^a_{\mu \nu} = \partial_\mu A^a_\nu - \partial_\nu A^a_\mu + g f^{abc} A^b_\mu A^c_\nu$
is the field strength; $a, b,c,d =$ either $1, 2, 3$ or $2,5,7$ are the $SU(2)$ colour indices; $m,n = $ either $4,5, \cdots , 8$ or $1,3,4,6,8$; $g$ 
is the coupling constant; $f^{ABC}$ are the structure constants for the $SU(3)$ gauge group; $A, B, C = 1,2, \cdots , 8$. The equation \eqref{5-10} 
describes $SU(2) \in SU(3)$ degrees of freedom that have non-zero expectation values, and equation \eqref{5-20} describes coset $SU(3) / SU(2)$ 
degrees of freedom with zero expectation values:
\begin{eqnarray}
	\hat A^{a \mu} &=& \left\langle \hat A^{a \mu} \right\rangle +
	i \delta \hat A^{a \mu} ,
\label{5-90}\\
	\left\langle \hat A^{m \mu} \right\rangle &=& 0 .
\label{5-100}
\end{eqnarray}
We seek a cylindrically symmetric solution of equations  \eqref{5-10} and \eqref{5-20} in the subgroup $SU(2) \in SU(3)$ spanned on either $\lambda^{1,2,3}$ or $\lambda^{2,5,7}$ in the form
\begin{eqnarray}
	A^{1,2}_t(\rho) &=& \frac{f(\rho)}{g} , \quad
	A^{1,2}_z(\rho) = \frac{u(\rho)}{g} ,
\label{1-110}\\
	A^{2,5}_t(\rho) &=& \frac{w(\rho)}{g} , \quad
	A^{2,5}_z(\rho) = \frac{v(\rho)}{g} ,
\label{5-120}\\
	\phi(\rho) &=& \frac{\phi(\rho)}{g}.
\label{5-130}
\end{eqnarray}
Here the first superscript indices are for $\lambda^{1,2,3}$ and the second ones -- for $\lambda^{2,5,7}$. We work in a cylindrical coordinate 
system $z, \rho, \varphi$, and the corresponding colour electric and magnetic fields are then
\begin{eqnarray}
	E^{3,7}_z(\rho) &=& F^{3,7}_{tz} = \left( E^{3,7}_z \right)_1 - \left( E^{3,7}_z \right)_2  =
	\frac{fv - w u}{g} ,
\label{5-140}\\
	E^{1,2}_\rho(\rho) &=& F^{1,2}_{t \rho} =
	- \frac{f'(\rho)}{g} , \quad
	E^{2,5}_\rho(\rho) = F^{2,5}_{t \rho} = - \frac{w'(\rho)}{g} ,
\label{5-150}\\
	H^{2,5}_\varphi (\rho) &=& \epsilon_{\varphi \rho z} F^{2,5 \; \rho z} =
	- \frac{v'(\rho)}{g} , \quad
	H^{1,2}_\varphi (\rho) = \epsilon_{\varphi \rho z} F^{1,2 \; \rho z} =
	- \frac{u'(\rho)}{g},
\label{5-160}
\end{eqnarray}
where
$\left( E^{3,7}_z \right)_1 = A^{1,2}_t A^{2,5}_z = \frac{fv}{g},
\left( E^{3,7}_z \right)_2 = A^{2,5}_tA^{1,2}_z = - \frac{w u }{g}$.
For simplicity, we consider the case with
\begin{equation}
	w = f, u = v.
\label{5-170}
\end{equation}
In both cases we have the following set of equations (for details see Appendix \ref{a1})
\begin{eqnarray}
	- f'' - \frac{f'}{\rho} + m^2 \phi^2 f &=& \mu^2 f ,
\label{5-180}\\
	-v'' - \frac{v'}{\rho} + m^2 \phi^2 v &=& \mu^2 v ,
\label{5-190}\\
	\phi'' + \frac{\phi'}{\rho} &=& \phi \left[
		\tilde \alpha \left(
			- f^2 + v^2
		\right) + \lambda \left(
			\phi^2 - M^2
		\right)
	\right] .
\label{5-200}
\end{eqnarray}
We see that equations \eqref{5-180} and \eqref{5-190} are Schr\" odinger-type equations with a solution
$v(\rho) = k f(\rho)$, where $k$ is a constant, $\phi$ is the potential and $\mu$ is an eigenvalue. In this case we can rewrite the set of equations \eqref{5-180}-\eqref{5-200} as follows
\begin{eqnarray}
	- f'' - \frac{f'}{x} + \phi^2 f &=& \mu^2 f ,
\label{5-210}\\
	\phi'' + \frac{\phi'}{x} &=& \phi \left[
	\alpha f^2 + \lambda \left( \phi^2 - M^2 \right)
	\right].
\label{5-220}
\end{eqnarray}
Here $\alpha = \tilde \alpha (k^2 - 1)$ and it can be an arbitrary real number; we redefined
$m \phi / f(0) \rightarrow \phi$, $\lambda /m^2 \rightarrow \lambda$, $m M \rightarrow M$,
$f / f(0) \rightarrow f$, $x = \rho f(0)$. Numerical investigation shows that regular solution to \eqref{5-210} and \eqref{5-220} does exist only for some positive $\alpha > 0$. 
It is necessary to note that because of  \eqref{5-170} the total longitudinal electric field 
$E^{3,7}_z = 0$,  and this leads to the fact that equations \eqref{5-180}, \eqref{5-190}, 
and \eqref{5-210} do not have non-linear terms like $f v^2$.

\begin{figure}[h!]
	\begin{minipage}[ht]{.45\linewidth}
		\begin{center}
			\fbox{
				\includegraphics[width=.9\linewidth]{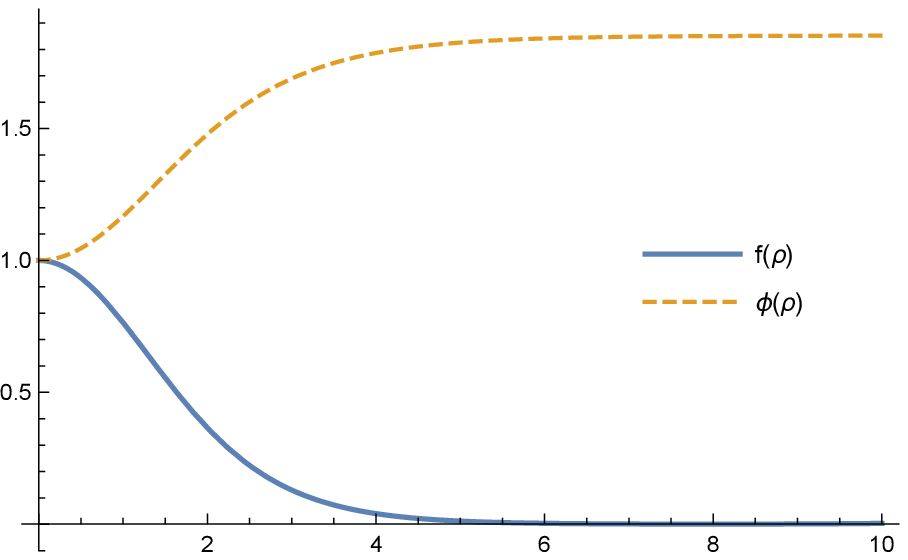}
			}
		\end{center}
		\caption{The profile of the functions $f(\rho), \phi(\rho)$.
		}
		\label{potentials_zeroFT}
	\end{minipage}
	\begin{minipage}[ht]{.45\linewidth}
		\begin{center}
			\fbox{
				\includegraphics[width=.9\linewidth]{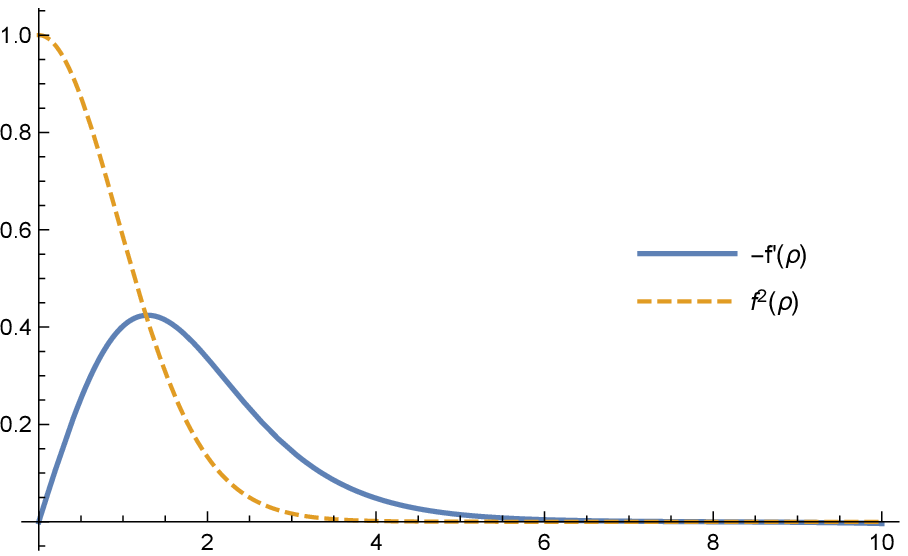}
			}
		\end{center}
		\caption{The profiles of electric and magnetic fields
			$g E^{1,2}_\rho(\rho) = g E^{2,5}_\rho(\rho) = - f'(\rho)$,
			$g H^{2,5}_\varphi (\rho) = g H^{1,2}_\varphi (\rho) = - f'(\rho)$,
			$g \left( E^{3,7}_z \right)_1 = - g \left( E^{3,7}_z \right)_2 = f^2(\rho)$.
		}
		\label{fields_zeroFT}
	\end{minipage}\hfill
\end{figure}

The results of numerical calculations are presented in Figs.~\ref{potentials_zeroFT} and \ref{fields_zeroFT}. We see that we have the dual Meissner effect: the longitudinal electric fields $\left( E^{3,7}_z \right)_1$ and
$\left( E^{3,7}_z \right)_2$ are confined into a tube by the scalar field $\phi$
(which is a condensate of the coset fields).

\section{NP quantization \`{a} la Heisenberg vs turbulence modelling}

In this section we would like to show that the two-equation approximation for the NP quantization \`{a} la Heisenberg and the Reynolds
approximation in stochastic turbulence modeling have the same mathematical basis:
the truncation of the corresponding infinite set of equations  for all either Green functions or cumulants.

The Navier-Stokes equation describing a turbulent flow  is (in this section we follow the textbook~\cite{Wilcox})
\begin{equation}
	\rho \left(
	\frac{\partial v_i}{\partial t}
	+ v_j \cdot \frac{\partial v_i}{\partial x_j}
	\right) = - \frac{\partial p}{\partial x_i} +
	\frac{\partial t_{ij}}{\partial x_j},
	\label{3-10}
\end{equation}
where $v_i$ is the flow velocity, $\rho$ is the fluid density, $p$ is the pressure, $t_{ij} = 2 \mu s_{ij}$ is the viscous stress tensor,
$s_{ij} = \frac{1}{2} \left(
\frac{\partial v_i}{\partial x_j} + \frac{\partial v_j}{\partial x_i}
\right)$, and $\mu$ is the molecular viscosity. All physical quantities are random ones and statistically fluctuate.

In order to show the similarity between the NP quantization and turbulence modelling, we will obtain two equations:
first of them will be the averaged Navier-Stokes equation, and the second one is obtained by multiplying the Navier-Stokes equation
by the velocity with subsequent averaging. The details can be found in the textbook \cite{Wilcox}, and the result is as follows:
\begin{eqnarray}
	\rho \frac{\partial V_i}{\partial t} +
	\rho V_j \frac{\partial V_i}{\partial x_j}
	&=& - \frac{\partial p}{\partial x_i} +
	\frac{\partial }{\partial x_j} \left(
	2 \mu S_{ji} - \overline{\rho v_j^\prime v_i^\prime}
	\right) ,
	\label{3-20}\\
	\frac{\partial \tau_{ij}}{\partial t} +
	V_k \frac{\partial \tau_{ij}}{\partial x_k} &=&
	- \tau_{ik} \frac{\partial V_j}{\partial x_k} -
	\tau_{jk} \frac{\partial V_i}{\partial x_k} +
	2 \mu \overline{\frac{\partial v^\prime_i}{\partial x_k}
		\frac{\partial v^\prime_j}{\partial x_k}} -
	\overline{
		p^\prime \left(
		\frac{\partial v_i^\prime}{\partial x_j} +
		\frac{\partial v_j^\prime}{\partial x_i}
		\right)
	} +
	\frac{\partial}{\partial x_k} \left(
	\nu \frac{\partial \tau_{ij}}{\partial x_k} +
	C_{ijk}
	\right),
	\label{3-30}
\end{eqnarray}
where $v_i(\vec x, t)$ is the instantaneous velocity; $V_i(\vec x, t)$ is a mean value of velocity; $v^\prime_i(\vec x, t)$ is a fluctuating part of the velocity;
$\tau_{ij} = - \overline{\rho v_i^\prime v_j^\prime}$ is the Reynolds stress tensor;
$\overline{( \cdots )}$ is the statistical averaging;
$
C_{ijk} =
\overline{\rho v^\prime_i \rho v^\prime_j \rho v^\prime_k} +
\delta_{jk} \overline{p^\prime v^\prime_i} +
\delta_{ik} \overline{p^\prime v^\prime_j}
$. Two equations \eqref{3-20} and \eqref{3-30} are not closed, since we have  new unknown functions
$\overline{\rho v^\prime_i \rho v^\prime_j \rho v^\prime_k}$, $\overline{\mu \frac{\partial v^\prime_i}{\partial x_k}
	\frac{\partial v^\prime_j}{\partial x_k}}$, and
$\overline{v^\prime_i \frac{\partial p^\prime}{\partial x_j}}$ . The equation \eqref{3-20} is the averaged Navier-Stokes equation, and \eqref{3-30} is the Reynolds equation. In order to close \eqref{3-20} and \eqref{3-30}, we have to make some physically reasonable assumptions about the unknowns, namely $C_{ijk}$,
$\overline{\rho v^\prime_i \rho v^\prime_j \rho v^\prime_k}$, $\overline{\mu \frac{\partial v^\prime_i}{\partial x_k}
	\frac{\partial v^\prime_j}{\partial x_k}}$, and
$\overline{v^\prime_i \frac{\partial p^\prime}{\partial x_j}}$.

We can compare the NP quantization and turbulence modelling in the following table:

\fbox{
	\addtolength{\linewidth}{-1\fboxsep}%
	\addtolength{\linewidth}{-1\fboxrule}%
	\begin{minipage}{\linewidth}
		\begin{eqnarray*}
			\text{quantum averaged Yang - Mills} &\leftrightarrow &
			\text{stochastically averaged Navier-Stokes eqn.},
			\\
			\text{quantum averaged $A^m_\mu \cdot $ Yang - Mills} &\leftrightarrow &
			\text{stochastically averaged Reynolds eqn., }
			\\	
			\text{closure problem with a 4-point Green function } G^{(4)} &\leftrightarrow &
			\text{closure problem with } C_{ijk},
			\overline{\rho v^\prime_i \rho v^\prime_j \rho v^\prime_k},
			\overline{\mu \frac{\partial v^\prime_i}{\partial x_k}
				\frac{\partial v^\prime_j}{\partial x_k}},
			\overline{v^\prime_i \frac{\partial p^\prime}{\partial x_j}} .
		\end{eqnarray*}
	\end{minipage}
}

\vspace{4mm}
The table gives us a full understanding of close interrelation between the NP quantization and turbulence modeling.
In addition,  we want to note that the diagram technique from perturbative quantum field theory can be applied in turbulence modeling, for details see the textbook \cite{mccomb}.

\section{Discussion and conclusions}

The physical meaning of the two-equation approximation is to describe physical systems in which one
group of degrees of freedom is practically in a classical phase, and the remaining group of degrees of freedom is in a pure quantum phase.
In addition, in the first group, we have quantum fluctuations around the mean values.
The dispersion of these fluctuations gives rise to the appearance of masses of the corresponding gauge fields.
The dispersion of quantum fluctuations in the second group gives rise to the gluon condensate.
In fact, this system is a system where classical non-Abelian gauge fields belonging to a subgroup interact
with the quantum condensate of gauge fields belonging to the coset.
The most interesting case here can be the case when classical non-Abelian gauge fields are confined by a condensate of quantum gauge fields.
For example, it can be: (a) a flux tube with a longitudinal color electric field stretched between infinitely separated quark and antiquark;
(b) a hedgehog with an exponentially decreasing non-Abelian gauge field which is pushed out by the condensate, or something like this.

We have shown that applying the two-equation approximation in the non-perturbative quantization \`{a} la Heisenberg for QCD one can obtain the flux tube stretched between: (a) quark and antiquark, (b) quark and quark, and (c) antiquark and antiquark located at $\pm \infty$ with non-zero and zero longitudinal color electric field. It is shown that all color electric and magnetic fields are expelled by the scalar field that describes a condensate of coset non-Abelian fields. This effect is the analog of the Meissner effect in superconductivity for non-Abelian color fields.

One of the problems in QCD is to show that a flux tube filled with a longitudinal electric field does appear between quark and antiquark.
The conventional opinion in this case is that the appearance of the flux tube is the manifestation of the dual Meissner effect --
the pushing out of color electric field from the gluon condensate. We have shown that in our two-equation approximation such a solution does really exist.
That means that our approach can actually describe the dual Meissner effect.

As we mentioned above, probably the simplest way to close two equations \eqref{1-160} and \eqref{1-170}
is the Mexican hat approximation \eqref{1-180}. But one can assume that there are many other possibilities.
For example, Green functions of the operator Yang-Mills equation \eqref{1-10} can be connected with solutions in some classical field theory.
For instance, one can investigate a holographic idea that QCD Green functions are connected with a classical action of some classical theory calculated on a hypersurface of a bulk.

It would be interesting to compare Green functions calculated in this approach with results obtained in lattice calculations, see Ref's \cite{D'Elia:2015dxa} - \cite{D'Elia:2002ck}.

\section*{Acknowledgements}

This work was supported by Grant 3101/GF4 IPC-11 in fundamental research in natural sciences by the  Ministry Education ans Science of Republic of Kazakhstan. I am very grateful to N. Kochelev and V. Folomeev for fruitful discussions and comments.

\appendix
\section{Quantum averaged equations}
\label{averaged}

In this appendix we would like to give detailed calculations of first two equations \eqref{1-20} and \eqref{1-30} for the two-equation approximation. The equations with $A = a, m$ are
\begin{eqnarray}
	D_\nu \hat F^{a \mu \nu} &=& \tilde D_\nu \hat{\mathcal F}^{a \mu \nu} +
	g f^{amn} \left[ \partial_\nu \left(
			\hat A^{m \mu} \hat A^{n \nu}
		\right) +
		\hat A^m_\nu \hat \partial^{[\mu} \hat A^{(a) \nu]}
	\right] +
\nonumber \\
	&&
		g^2 \left( f^{abc} f^{cpq} \hat A^b_\nu \hat A^{p\mu} \hat A^{q\nu} -
		f^{amn} f^{bnp} \hat A^m_\nu \hat A^{b [\mu} \hat A^{(p) \nu]}
	\right) = 0 ,
\label{a-10}\\
	D_\nu \hat F^{m \mu \nu} &=& \partial_\nu \left(
		\partial^{[\mu} \hat A^{(m) \nu}
	\right) + g f^{amn} \left[
		- \partial_\nu A^{a[\mu} \hat A^{(n) \nu]} -
		\hat A^a_\nu \partial^{[\mu} A^{(n) \nu]} +
		\hat A^n_\nu \hat{\mathcal F}^{a \mu \nu}
	\right] +
\nonumber \\
	&&	g^2 f^{amn} \left(
		f^{bnp} \hat A^a_\nu \hat A^{b[\mu} \hat A^{(p) \nu]} +
		f^{apq} \hat A^n_\nu \hat A^{p \mu} \hat A^{q \nu}
	\right) = 0.
\label{a-20}
\end{eqnarray}
Here $\tilde D_\nu = \partial_\nu + f^{abc} \hat A^b_\nu$ is the covariant derivative in the subgroup $\mathcal G \subset SU(N)$;
$\hat A^{p [\mu} \hat A^{(q) \nu]} = \hat A^{p \mu} \hat A^{q \nu} -
\hat A^{p \nu} \hat A^{q \mu}
$ is the antisimmetrization over $\mu$ and $\nu$;
$\hat{\mathcal F}^{a \mu \nu} = \partial_\mu \hat A^a_\nu - \partial_\nu \hat A^a_\mu +
f^{abc} \hat A^b_\mu \hat A^c_\nu
$; $\hat A^a_\mu = A^a_\mu + i \delta \hat A^a_\mu$; the expectation value of
$\hat A^a_\mu$ is $\left\langle \hat A^a_\mu \right\rangle = A^a_\mu$.

2-point Green function
$G^{ab \mu \nu} = \left\langle \delta \hat A^{a\mu}(y) \hat \delta A^{b\nu(x)} \right\rangle$ describes the dispersion defect in the following manner. Let us consider
\begin{eqnarray}
	\left\langle
		\left( \hat A^a_\mu \right)^\dagger \hat A^b_\nu
	\right\rangle &=& A^a_\mu A^b_\nu + \left\langle
	\delta \hat A^a_\mu \delta \hat A^b_\nu
	\right\rangle ,
\label{a-24}\\
	\left\langle
		\hat A^a_\mu \hat A^b_\nu
	\right\rangle &=& A^a_\mu A^b_\nu - \left\langle
	\delta \hat A^a_\mu \delta \hat A^b_\nu
	\right\rangle	.
\label{a-28}
\end{eqnarray}
Here
$
\left( \hat A^a_\mu \right)^\dagger = A^a_\mu - i \delta \hat A^a_\mu
$;
$
\left( \delta \hat A^a_\mu \right)^\dagger = \delta \hat A^a_\mu
$. We see that
$
\left\langle
\hat A^a_\mu \hat A^b_\nu \right\rangle <
\left\langle \hat A^a_\mu \right\rangle \left\langle \hat A^b_\nu \right\rangle
$ that means that
$
\left\langle
\delta \hat A^a_\mu \delta \hat A^b_\nu
\right\rangle	
$ can be referred as the dispersion defect.

Quantum averaging of \eqref{a-10} gives us the following equation:
\begin{equation}
\begin{split}
	\left\langle D_\nu F^{a \mu \nu}  \right\rangle = &
	\tilde D_\nu \mathcal F^{a \mu \nu} +
	g f^{amn} \left\{
		\partial_{x^\nu} G^{mn \mu \nu}(x,x) + \left[
			\partial^{x^\mu} G^{mn \phantom{\nu} \nu}_{\phantom{mn} \nu}(y,x) -
			\partial^{x^\nu} G^{mn \phantom{\nu} \mu}_{\phantom{mn} \nu}(y,x)
		\right]_{y=x}
	\right\} +
\\
	&
	g^2 \left[ f^{abc} f^{cpq} A^b_\nu G^{pq \mu \nu} -
	f^{amn} f^{bnp} \left(
		A^{b\mu} G^{mp \phantom{\nu} \nu}_{\phantom{mp} \nu} -
		A^{b\nu} G^{mp \phantom{\nu} \mu}_{\phantom{mp} \nu}
	\right)
	\right] +
\\
	&
	\left( \mu^2 \right)^{ab \mu \nu} A^b_\nu = 0,
\label{a-30}
\end{split}
\end{equation}
where
\begin{eqnarray}
	\left\langle
		\tilde D_\nu \hat{\mathcal F}^{a \mu \nu}
	\right\rangle &=& \tilde D_\nu \mathcal F^{a \mu \nu} +
	\left( \mu^2 \right)^{ab \mu \nu} A^b_\nu ,
\label{a-35} \\
	\left( \mu^2 \right)^{ab \mu \nu} &=& - g^2 \left(
		f^{abc} f^{cde} G^{de \mu \nu} +
		\eta^{\mu \nu} f^{adc} f^{cbe} G^{de \phantom{\alpha} \alpha}_{\phantom{de} \alpha} +
		f^{aec} f^{cdb} G^{ed \nu \mu}
	\right) ,
\label{a-40}
\end{eqnarray}
$G^{ab \mu \nu}(y,x) = \left\langle \delta \hat A^{a\mu}(y) \hat \delta A^{b\nu(x)} \right\rangle$ and
$G^{mn \mu \nu}(y,x) = \left\langle \hat A^{m\mu}(y) \hat A^{n\nu}(x) \right\rangle$ are 2-point Green functions.
In deriving \eqref{a-30}, we took into account that we consider a physical system where quantum degrees
of freedom from the subgroup $\mathcal G$ and the coset $SU(N) / \mathcal G$ do not correlate:
\begin{equation}
	\left\langle
		\delta \hat A^{a_1}_{\mu_1}(x) \ldots \hat A^{m_1}_{\nu_1}(x) \ldots
	\right\rangle \approx 0 .
\label{a-50}
\end{equation}
The averaging of the equation \eqref{a-20} gives us
\begin{equation}
	\left\langle D_\nu \hat F^{m \mu \nu}  \right\rangle = 0,
\label{a-80}
\end{equation}
since $\left\langle \hat A^m_\mu  \right\rangle =
\left\langle \hat A^m_\mu \hat A^n_\nu \hat A^p_\alpha \right\rangle = 0
$, because of the restriction \eqref{1-80}.

The multiplication of \eqref{a-20} by $\hat A^{r \alpha}$, followed by the quantum averaging, gives us
\begin{equation}
\begin{split}
	& \left\langle
		\hat A^{r \alpha} D_\nu \hat F^{m \mu \nu}
	\right\rangle = \left[
		\partial_{x^\nu} \partial^{x^\mu} G^{rm \alpha \nu} (y, x) -
		\Box_x G^{rm \alpha \mu} (y, x)
	\right]_{y = x} +
\\
	&
	g f^{amn} \left\lbrace
		- \partial_{x^\nu} \left[
			A^{a \mu} (x) G^{rn \alpha \nu} (y, x) -
			A^{a \nu} (x) G^{rn \alpha \mu} (y, x)
		\right]_{y = x} -
		A^a_\nu (x) \left[
			\partial^{x^\mu} G^{rn \alpha \nu} (y, x) -
			\partial^{x^\nu} G^{rn \alpha \mu} (y, x)
		\right]_{y = x} +
	\right.
\\
	&		
	\left.
		G^{rn \alpha}_{\phantom{rn \alpha} \nu} (x, x)
		\mathcal F^{a \mu \nu}(x)
	\right\rbrace +
	g^2 f^{amn} f^{bnp}  A^a_\nu (x)
		\left(
			A^{b \mu} (x) G^{rp \alpha \nu} (x, x) - A^{b \nu} (x) G^{rp \alpha \mu} (x, x)
		\right) +
\\
	&
	g^2 f^{amn} f^{apq} G^{rnpq \alpha \phantom{\nu} \mu \nu}_{\phantom{rnpq \alpha} \nu} (x,x,x,x)
	= 0.
\label{a-90}
\end{split}
\end{equation}
Here we took into account that quantum degrees of freedom from the subgroup $\mathcal G$ and the coset
$SU(N) / \mathcal G$ do not correlate,
$
\left\langle
	\delta \hat A^a_\nu \delta \hat A^{b \mu} \hat A^{p \nu} \hat A^{r \alpha}
\right\rangle \approx 0
$, and
\begin{equation}
	G^{rnpq \alpha \phantom{\nu} \mu \nu}_{\phantom{rnpq \alpha} \nu} (x,x,x,x) =
	\left\langle
		\hat A^{r \alpha} (x) \hat A^n_\nu (x) \hat A^{p \mu} (x) \hat A^{q \nu} (x)
	\right\rangle
\label{a-100}
\end{equation}
is the 4-point Green function that, following to the closure procedure, should be expressed in terms of 2-point Green functions.

The multiplication of \eqref{a-20} by $\hat A^c_\alpha$ with subsequent quantum averaging gives us
\begin{equation}
	\left\langle
		\hat A^c_\alpha(x) D_\nu \hat F^{m \mu \nu}(y) 
	\right\rangle = 0,
\label{a-110}
\end{equation}
since all quantum degrees of freedom from the subgroup $\mathcal G$ and the coset
$SU(N) / \mathcal G$ do not correlate.

Finally, we have the following two-equation approximation for the NP quantization \`{a} la Heisenberg for non-Abelian gauge theory in the static case with the limitations \eqref{1-80} and \eqref{a-50}:
\begin{eqnarray}
	&&
	\tilde D_\nu \mathcal F^{a \mu \nu} - \left[
		\left( m^2 \right)^{ab \mu \nu} -
		\left( \mu^2 \right)^{ab \mu \nu}
	\right] A^b_\nu = j^{a \mu} ,
\label{a-120} \\
	&&
	\left[
	\partial_{x^\nu} \partial^{x^\mu} G^{rm \alpha \nu} (y, x) -
	\Box_x G^{rm \alpha \mu} (y, x)
	\right]_{y = x} +
\nonumber \\
	&&
	g f^{amn} \biggl \{
	- \partial_{x^\nu} \left[
	A^{a \mu} (x) G^{rn \alpha \nu} (y, x) -
	A^{a \nu} (x) G^{rn \alpha \mu} (y, x)
	\right]_{y = x} -
	A^a_\nu (x) \left[
	\partial^{x^\mu} G^{rn \alpha \nu} (y, x) -
	\partial^{x^\nu} G^{rn \alpha \mu} (y, x)
	\right]_{y = x} +
\nonumber \\
	&&		
	G^{rn \alpha}_{\phantom{rn \alpha} \nu} (x, x)
	\mathcal F^{a \mu \nu}(x)
	\biggl \} +
	g^2 f^{amn} f^{bnp}  A^a_\nu (x)
	\biggl[
	A^{b \mu} (x) G^{rp \alpha \nu} (x, x) - A^{b \nu} (x) G^{rp \alpha \mu} (x, x)
	\biggl] +
\nonumber \\
	&&
	g^2 f^{amn} f^{apq} G^{rnpq \alpha \phantom{\nu} \mu \nu}_{\phantom{rnpq \alpha} \nu} (x,x,x,x)
	= 0,
\label{a-130}
\end{eqnarray}
where
\begin{eqnarray}
	\left( m^2 \right)^{ab \mu\nu} &=& - g^2 \left[
		f^{abc} f^{cpq} G^{pq \mu\nu} -
		f^{amn} f^{bnp} \left(
			\eta^{\mu \nu} G^{mp \phantom{\alpha} \alpha}_{\phantom{mn} \alpha} -
			G^{mp \nu \mu}
		\right)
	\right] ,
\label{a-132} \\
	j^{a \mu} &=& - g f^{amn} \left\{
	\partial_{x^\nu} G^{mn \mu \nu}(x,x) + \left[
	\partial^{x^\mu} G^{mn \phantom{\nu} \nu}_{\phantom{mn} \nu}(y,x) -
	\partial^{x^\nu} G^{mn \phantom{\nu} \mu}_{\phantom{mn} \nu}(y,x)
	\right]_{y=x}
	\right\}.
\label{a-134}
\end{eqnarray}
The equations \eqref{a-30}, \eqref{a-90}, \eqref{a-80}, and \eqref{a-110} are correct with the approximations \eqref{1-80} and \eqref{a-50}.

\section{Scalar approximation for the condensate equation}
\label{approximation}

The equation \eqref{a-130} is an exact description of field quantum fluctuations. Generally speaking,  2-point Green functions
for different indices  have different behavior. In this section we want to consider a situation when all 2-point Green functions
for coset degrees of freedom have the same order and are approximately defined by the sum of all dispersions. In this case we can assume
that every 2-point Green function is proportional to the dispersion
\begin{equation}
	G^{mn \mu \nu}(y, x) \approx C^{mn \mu \nu} D(y, x),
\label{3c-10}
\end{equation}
where $C^{mm \mu}_{\phantom{mn\mu}\mu}$ is a constant. $D(x,x)$ is the sum of quantum field dispersions $\hat A^m_\mu$ when $y = x$.

In order to obtain equation for the total dispersion $D(x,x)$, we sum \eqref{a-130} over $r,m$ and $\alpha, \mu$.
To make the equation for the dispersion $D$ as simple as possible, we use the following $\mathfrak{Ansatz}$
\begin{equation}
	C^{mn \mu \nu} = \delta^{mn} \mathcal A^\mu \mathcal A^\nu,
\label{3c-30}
\end{equation}
where $\delta^{mn}$ is the Kronecker symbol; $\mathcal A^\mu \mathcal A_\mu$ is a constant;
$
	\mathcal A^\mu \partial_\mu = \partial_\mu \mathcal A^\mu = 0
$. The system of equations  \eqref{a-120} and \eqref{a-130} is not closed since we have the 4-point Green function $G^{mnpq \alpha \beta \mu \nu}$.
In order to close this set of equations,  we have to make some assumption about a 4-point Green function, namely, how it is connected with a 2-point Green function.
As mentioned above, we assume that a 4-point Green function is a bilinear combination of 2-point Green functions:
\begin{equation}
	G^{mnpq \alpha \beta \mu \nu}(x,x,x,x) \approx C^{mnpq \alpha \beta \mu \nu}
	D(x,x) \left[ M^2 - D(x,x) \right],
\label{3c-40}
\end{equation}
where $C^{mnpq \alpha \beta \mu \nu}$ and $M^2$ are constants. After that point,
we obtain the following equation for the total dispersion of the coset quantum fields $\hat A^m_\mu$:
\begin{equation}
\begin{split}
	& \Box_x D(y,x )_{y=x} -
	\left( m^2_\phi \right)^{ab \mu \nu} A^a_\nu(x) A^b_\mu(x) D(x,x) -
	\lambda D(x,x) \left[
	M^2 - D(x,x)
	\right] = 0,
\label{3c-50}
\end{split}
\end{equation}
where
\begin{eqnarray}
	\left( m^2_\phi \right)^{ab \mu \nu} &=& 
	g^2 f^{amn} f^{bnp} \frac{
		G^{mp \mu \nu} - \eta^{\mu \nu} G^{mp \alpha}_{\phantom{mp \alpha} \alpha}
	}
	{G^{mm \alpha}_{\phantom{mm\alpha} \alpha}}	 = 
	g^2 f^{amn} f^{bnm}
	\frac{\mathcal A^\mu \mathcal A^\nu -
	\eta^{\mu \nu} \mathcal A^\alpha \mathcal A_\alpha}{3 \mathcal A^\alpha \mathcal A_\alpha} ,
\label{b-10}\\
	\lambda &=& g^2 f^{amn} f^{apq}
	\frac{C^{mnpq \phantom{\mu \nu} \mu \nu}_{\phantom{mnpq} \mu \nu}}
	{3 \mathcal A^\alpha \mathcal A_\alpha}.
\label{b-30}
\end{eqnarray}
The solution of this equation is sought in the form
\begin{equation}
	D(y, x) = \phi(y) \phi(x) .
\label{3c-60}
\end{equation}
We think that such an approximation can be used only \emph{in a static case for the description of some regular static objects}.
For example, it can be a hedgehog (a spherical object filled with radial electric/magnetic fields), or a flux tube in a gluon condensate, or a glueball and so on.

Upon inserting  \eqref{3c-60} into \eqref{3c-50}, we have the following equation
\begin{equation}
	\Box \phi -
	\left( m^2_\phi \right)^{ab \mu \nu} A^a_\nu A^b_\mu \phi -
	\lambda \phi \left(
	M^2 - \phi^2
	\right) = 0.
\label{3c-70}
\end{equation}
Physically this equation describes the gluon condensate $\phi$ formed by the coset quantum gauge fields
$A^m_\mu \in SU(N) / \mathcal G$ and interacting with the gauge fields
$A^a_\mu \in \mathcal G \subset SU(N)$.

\section{Evaluation of $\left( \mu^2 \right)^{ab\mu\nu}$, $\left( m^2 \right)^{ab\mu\nu}$, and
	$\left( m^2_\phi \right)^{ab\mu\nu}$for the flux tube solution}
\label{adaptation}

The equation for this case is
\begin{equation}
	\tilde D_\nu \mathcal F^{a \mu \nu} - \left[
	\left( m^2 \right)^{ab \mu \nu} -
	\left( \mu^2 \right)^{ab \mu \nu}
	\right] A^b_\nu = j^{a \mu} .
\label{c-10}
\end{equation}
In order to obtain a flux tube solution, we need to consider a  physical system with some special properties.
We will consider $\mathcal G = SU(2) \times U(1)$, $SU(N) = SU(3)$ and a physical system where 2-point Green functions  can approximately be expressed as follows:
\begin{equation}
	G^{ab \mu \nu} (y,x) \approx \Delta^{ab}
	\mathcal B^\mu \mathcal B^\nu,
\label{c-20}
\end{equation}
where $\mathcal B_\mu \mathcal B^\mu$ is a constant and
\begin{equation}
	\Delta^{ab} = \begin{pmatrix}
		\delta_1	& 0			&  0 	\\
		0			& \delta_2	&  0	\\
		0			& 0			&  \delta_3
	\end{pmatrix}.
\label{c-30}
\end{equation}
This equation shows us that we have some color anisotropy in the $SU(2)$ subgroup in the sense that
$
G^{11\mu\nu} \neq G^{22\mu\nu} \neq G^{33\mu\nu}
$. We choose the vectors $\mathcal A^\mu$ and $\mathcal B^\mu$  in the form
\begin{eqnarray}
	\mathcal A^\mu = \left( 0, 0, 0, \frac{\mathcal A_\varphi}{\rho} \right) ,
\label{c-40}\\
	\mathcal B^\mu = \left( 0, 0, \mathcal B_\rho, \frac{\mathcal B_\varphi}{\rho} \right).
\label{c-50}
\end{eqnarray}
Here we work in the cylindrical coordinate system
$
ds^2 = dt^2 - dz^2 - d \rho^2 - \rho^2 d \varphi^2
$. We seek the vector potential for the flux tube solution  in the form
\begin{equation}
	A^1_t \neq 0 ; \quad A^2_z \neq 0 .
\label{c-70}
\end{equation}
With such choice of the vectors $A^a_\mu$, $\mathcal B^\mu$ and $\mathcal A^\mu$, we have
\begin{eqnarray}
	\left( \mu^2 \right)^{1b t\nu} A^b_\nu =
	g^2 \left( \mathcal B_\rho^2 + \mathcal B_\varphi^2 \right)
	\left( \delta_2 + \delta_3 \right) A^1_t &=& \mu_1^2 A^1_t,
\label{c-80}\\
	\left( \mu^2 \right)^{2b z\nu} A^b_\nu =
	- g^2 \left( \mathcal B_\rho^2 + \mathcal B_\varphi^2 \right)
	\left( \delta_1 + \delta_3 \right) A^2_z &=& - \mu_2^2 A^2_z,
\label{c-90}\\
	\left( m^2 \right)^{1b t\nu} A^b_\nu =
	\left( 3 g^2 \mathcal A_\varphi^2 \right) \phi^2 A^1_t &=& m^2 \phi^2 A^1_t,
\label{c-60}\\
	\left( m^2 \right)^{2b z\nu} A^b_\nu =
	- \left( 3 g^2 \mathcal A_\varphi^2 \right) \phi^2 A^2_z &=& - m^2 \phi^2 A^2_z,
\label{c-100}\\
	\left( m^2_\phi \right)^{ab \mu \nu} A^a_\nu A^b_\mu =
	\frac{g^2}{4} \left[
		\left( A^1_t \right)^2 - \left( A^2_z \right)^2
	\right] &=& m^2_\phi \left[
	\left( A^1_t \right)^2 - \left( A^2_z \right)^2
	\right] .
\label{c-110}
\end{eqnarray}
Here $\mu_2 \neq \mu_1$ if $\delta_2 \neq \delta_1$. Insertion of \eqref{3c-10} and \eqref{3c-30} into \eqref{a-134} yields
\begin{equation}
	j^{a\mu} = 0 .
\label{c-120}
\end{equation}

\section{Coefficients of equations \eqref{5-180}-\eqref{5-200} for flux tube with zero longitudinal electric field}
\label{a1}

\subsection{$SU(2)$ subgroup spanned on $\lambda^{1,2,3}$}

We use the following \textgoth{Ans\"atze} for the 2-point Green functions $G^{ab \mu \nu}$
\begin{eqnarray}
	G^{ab \mu \nu} (y,x) &\approx& \Delta^{ab}
	\mathcal B^\mu \mathcal B^\nu ,\quad  a, b = 1, 2, 3 ;
\label{a1-10}\\
	C^{mn \mu \nu} &\approx& \delta^{mn} \mathcal A^\mu \mathcal A^\nu \phi^2 ,
	\quad  m, n = 4, 5, 6, 7,
\label{a1-20}
\end{eqnarray}
where $\mathcal B_\mu \mathcal B^\mu$ and $\mathcal A_\mu \mathcal A^\mu$ are constants and
\begin{eqnarray}
	\Delta^{ab} &=& \text{diag} \biggl(
		\delta_1, \delta_2, \delta_3, 0,0,0,0,0
	\biggl) ,
\label{a1-30}\\
	\delta^{mn} &=& \text{diag} \biggl(
		0,0,0,\Delta_4, \Delta_5, \Delta_6, \Delta_7, 0
	\biggl) ,
\label{a1-40}\\
	\mathcal A^\mu &=& \left( 0, 0, \mathcal A_\rho, \frac{\mathcal A_\varphi}{\rho} \right) ,
\label{a1-50}\\
	\mathcal B^\mu &=& \left( 0, 0, \mathcal B_\rho, \frac{\mathcal B_\varphi}{\rho} \right).
\label{a1-60}
\end{eqnarray}
We choose $\mathcal B^\mu$ and  $\mathcal A^\mu$ in such form because $A^{\cdots}_{t,z}$ from \eqref{1-110} and \eqref{1-120} are non-zero: $A^{\cdots}_{t,z} \neq 0$. Substitution of \eqref{a1-30}-\eqref{a1-60} in \eqref{a1-10} and \eqref{a1-20} gives us [for $u(\rho) = v(\rho), w(\rho) = f(\rho)$]:
\begin{eqnarray}
	\mu_1^2 &=&	g^2 \left( \mathcal B_\rho^2 + \mathcal B_\varphi^2 \right)
	\left( \delta_2 + \delta_3 \right) ,
\label{a1-70}\\
	\mu_2^2 &=& g^2 \left( \mathcal B_\rho^2 + \mathcal B_\varphi^2 \right)
	\left( \delta_1 + \delta_3 \right) ,
\label{a1-80}\\
	m^2 &=& \frac{3}{4} g^2 \left(
		\mathcal A_\rho^2 + \mathcal A_\varphi^2
	\right) \left(
		\Delta_4 + \Delta_5 + \Delta_6 + \Delta_7
	\right) \phi^2 ,
\label{a1-90}\\
	\left( m^2_\phi \right)^{ab \mu \nu} A^a_\nu A^b_\mu &=&
	\frac{g^2}{2} \left( f^2 - v^2 \right) .
\label{a1-100}
\end{eqnarray}
We set $\delta_2 = \delta_1$ and then
\begin{equation}
	\mu_2^2 = \mu_1^2 .
\label{a1-110}
\end{equation}

\subsection{$SU(2)$ subgroup spanned on $\lambda^{2,5,7}$}

Similar construction can be done for the $SU(2)$ group spanned on $\lambda^{2,5,7}$ :
\begin{eqnarray}
	\Delta^{ab} &=& \text{diag} \biggl(
		0, \delta_2, 0, 0, \delta_5, 0, \delta_7, 0
	\biggl) ,
\label{b1-10}\\
	\delta^{mn} &=& \text{diag} \biggl(
		\Delta_1, 0, \Delta_3, \Delta_4,0,\Delta_6,0,\Delta_8
	\biggl) ,
\label{b1-20}\\
	\mathcal A^\mu &=& \left( 0, 0, \mathcal A_\rho, \frac{\mathcal A_\varphi}{\rho} \right) ,
\label{b1-30}\\
	\mathcal B^\mu &=& \left( 0, 0, \mathcal B_\rho, \frac{\mathcal B_\varphi}{\rho} \right).
\label{b1-40}
\end{eqnarray}
Substitution of \eqref{b1-10}-\eqref{b1-40} in \eqref{a1-10} and \eqref{a1-20} gives us [for $u(\rho) = v(\rho), w(\rho) = f(\rho)$]:
\begin{eqnarray}
	\mu_1^2 &=&	\frac{g^2}{4} \left( \mathcal B_\rho^2 + \mathcal B_\varphi^2 \right)
	\left( \delta_5 + \delta_7 \right) ,
\label{b1-50}\\
	\mu_2^2 &=& \frac{g^2}{4} \left( \mathcal B_\rho^2 + \mathcal B_\varphi^2 \right)
	\left( \delta_2 + \delta_7 \right) ,
\label{b1-60}\\
	m_1^2 &=& \frac{3}{4} g^2 \left(
		\mathcal A_\rho^2 + \mathcal A_\varphi^2
	\right) \left(
		4 \Delta_1 + 4 \Delta_3 + \Delta_4 + \Delta_6
	\right) ,
\label{b1-70}\\
	m_2^2 &=& \frac{3}{4} g^2 \left(
		\mathcal A_\rho^2 + \mathcal A_\varphi^2
	\right) \left(
		\Delta_1 + \Delta_3 + 4 \Delta_4 + \Delta_6 + 3 \Delta_8
	\right) ,
\label{b1-80}\\
	\left( m^2_\phi \right)^{ab \mu \nu} A^a_\nu A^b_\mu &=&
	\frac{g^2}{4} \frac{5 \Delta_1 + 5 \Delta_3 + 5 \Delta_4 + 2 \Delta_6 + 3 \Delta_8}
	{\Delta_1 + \Delta_3 + \Delta_4 + \Delta_6 + \Delta_8}
	\left( f^2 - v^2 \right) .
\end{eqnarray}
We set $\delta_5 = \delta_2$ and
$
4 \Delta_1 + 4 \Delta_3 + \Delta_4 + \Delta_6 =
\Delta_1 + \Delta_3 + 4 \Delta_4 + \Delta_6 + 3 \Delta_8
$ and then
\begin{equation}
	\mu_2^2 = \mu_2^2, \quad m_2^2 = m_1^2.
\label{b1-90}
\end{equation}

\end{document}